# Theoretical Calculation of the Electric Field in the Vicinity of a Pore Formed in a Cell Membrane

William A Hercules, James Lindesay, and Anna Coble
Howard University

**Abstract**

Electroporation of biological cell membranes is a phenomena that occurs almost universally at a transmembrane potential of approximately 1000 mV. Theories of pore formation have been proposed based on electrocompression (Crowley 1972, Malderelli and Steebe 1992 ), local membrane defects ( Abidor and Chizdhmazdev 1979, Weaver and Mintzer 1981, Sugar 1981), and  Statistical models  were proposed by (Sowers and Lieber 1986). However, these theories do not in general derive the form of the electric field within the pore. In this paper, we present a view that is useful in describing the biological membrane. We also derive an approximate functional form for the electric field in the region of a pore formed in the membrane.

**Introduction**

Biological cells exist in a wide variety of shapes and sizes, which is dependent on their function in the organism. Although, in general, the shape of a cell is determined by its special function in the organism, a great many of the cells are possessed of an amorphous shape. A consequence of the fluid-like nature of the cell membrane is its ability to adopt a different shape.  One of the most generally useful geometries that describe a wide range of cellular properties is that of a spherical shell. These cells may be partially embedded into a small pore in a material and made to assume an asymmetric dumbbell shape. Pores may be formed in the membrane by the application of an electric field of large enough magnitude across the membrane.  The cell membrane, which separates the interior of the cell from its surrounding, is composed of a lipid bilayer which is approximately 10 nm thick. This lipid bilayer is composed of long chain fatty acids with their hydrophobic ends apposed to form an interior region of the membrane, which possesses hydrophobic characteristics. The faces of the membrane are composed of hydrophilic polar groups.

The pores formed in the membrane of the cells may be transient, or permanent in nature, depending on the characteristics of the applied field.  For relatively low field strengths, the pores formed may be very small and short-lived. For intermediate field strengths, the pores may be larger and remain for longer times. For large applied fields the pores may be permanent.

A simple model of this arrangement is that of concentric spherical shells giving three regions. Region 1, has radius < $R_1$, region 2 which has inner radius $R_1$, and outer radius $R_2$, and region 3 which has radius > $R_2$. The three different regions of the shell namely region 1 with r < $R_1$, region 2 with $R_1$ < r < $R_2$ and region 3 with r > $R_2$, have conductivities given by $\sigma_1$, $\sigma_2$, $\sigma_3$ respectively. The potential distribution when these concentric shells are placed in a uniform electric field, can be readily found in spherical coordinates by solving Laplaces equation with appropriately chosen boundary conditions.

In order to determine the potential that exists in the region of a pore formed in the membrane, it is necessary to examine the cell at a closer level. This view corresponds to a point of view such that |r-$R_2$| << $R_2$, the radius of the cell. To a first approximation the cell membrane may be viewed as a flat sheet, lipid bi-layer. The poration of the membrane produces an opening in this sheet. We seek to determine the form of the electric field that exists in the region of the pore formed in the membrane by an application of a uniform external field. In this treatment, time dependent behaviors will not be considered. The external field is taken to be uniform in the z direction.

In spherical coordinates, the solution to Laplaces equation for the potential V, is

$$V(r,\theta) = \sum_l \left[ A_l\, r^l + B_l\, r^{-(l+1)} \right] P_l(\cos\theta)$$

where, due to the azimuthal symmetry and the requirement of periodicity of the $\Phi(\varphi)$ solution, we have set it to be a constant. The known boundary conditions may be expressed as

$$\text{For} \quad r \to \infty \quad V \Rightarrow -E_0\, z = -E_0\, r \cos\theta$$

$$\text{For} \quad r \to 0 \quad V \text{ is finite}$$

The potential in the three different regions may thus be written as

$$V_I(r,\theta) = \sum_l A_l \, r^l \, P_l(\cos\theta)$$

$$V_{II}(r,\theta) = \sum_l \left[ C_l \, r^l + D_l \, r^{-(l+1)} \right] P_l(\cos\theta)$$

$$V_{III}(r,\theta) = \sum_l \left[ F_l \, r^l + G_l \, r^{-(l+1)} \right] P_l(\cos\theta)$$

Inside the sphere includes $r = 0 \Rightarrow B = 0$ in region 1;

Outside the sphere, region 3, for which $r > R_2$, the potential must asymptotically approach the value, $-E_0 \, r \cos\theta$, thus the only values of l allowed is l=1.

At the interfaces between the regions, the fields must satisfy the two boundary conditions, namely

(I)     Tangential component of $\vec{E}$ is continuous,

(ii)     Normal $\vec{J}$ is continuous.

Taking $\sigma_1 \approx \sigma_3 \gg \sigma_2$ and writing $R_1 = R_2 - t$ where t is the membrane thickness, with $\dfrac{t}{R_2} \ll 1$. Application of these conditions, ignoring terms that are second order in small parameters, yields the coefficients as

$$A = \dfrac{3\,\sigma_2}{\sigma_1 \left[ 3\dfrac{\sigma_2}{\sigma_1} + 2\dfrac{t}{R_2} \right]} E_0 \qquad\qquad C = \dfrac{-\left(1 + 2\dfrac{\sigma_2}{\sigma_1}\right)}{\left[ 3\dfrac{\sigma_2}{\sigma_1} + 2\dfrac{t}{R_2} \right]} E_0$$

$$D = \dfrac{R_2^3 \left( 1 - 3\dfrac{t}{R_2} - \dfrac{\sigma_2}{\sigma_1} \right)}{\left[ 3\dfrac{\sigma_2}{\sigma_1} + 2\dfrac{t}{R_2} \right]} E_0 \qquad\qquad G = \left[ \dfrac{9\,R_2^3\,\sigma_2 - 18\,R_2^2\,t\,\sigma_2}{\left[ 9\,\sigma_2 + 4\dfrac{t}{R_2}\,\sigma_1 \right]} \right] E_0 - \dfrac{R_2^3}{2} E_0$$

and

$$F = E_0.$$

In addition to physically describing the membrane, the model used should also allow manipulable solutions of Laplace's equation. Of the various coordinate systems for which Laplace's equation is separable, the Oblate Spheroidal Coordinate system was chosen to be the most useful for modeling the membrane as well as the pore formed in the membrane.

**Oblate Spheroidal Coordinates**

In oblate spheroidal coordinates the transformation equations from Cartesian coordinates can be written as

$$x = a \cosh u \cos v \cos \varphi$$
$$y = a \cosh u \cos v \sin \varphi$$
$$z = a \sinh u \sin v.$$

Further, defining

$$\sinh u \Rightarrow \eta$$
$$\sin v \Rightarrow \zeta$$
$$\varphi \Rightarrow \varphi$$

yields the transformation equations

$$x = a \left(1 + \eta^2\right)^{\frac{1}{2}} \left(1 - \zeta^2\right)^{\frac{1}{2}} \cos \varphi$$

$$y = a \left(1 + \eta^2\right)^{\frac{1}{2}} \left(1 - \zeta^2\right)^{\frac{1}{2}} \sin \varphi$$

$$z = a \eta \zeta.$$

Calculating the metric scale factors yields

$$h_1 = h_\eta = a\left[\frac{(\eta^2 + \zeta^2)}{(1+\eta^2)}\right]^{\frac{1}{2}}$$

$$h_2 = h_\zeta = a\left[\frac{(\eta^2 + \zeta^2)}{(1-\zeta^2)}\right]^{\frac{1}{2}}$$

$$h_3 = h_\varphi = a\left[(1+\eta^2)(1-\zeta^2)\right]^{\frac{1}{2}}.$$

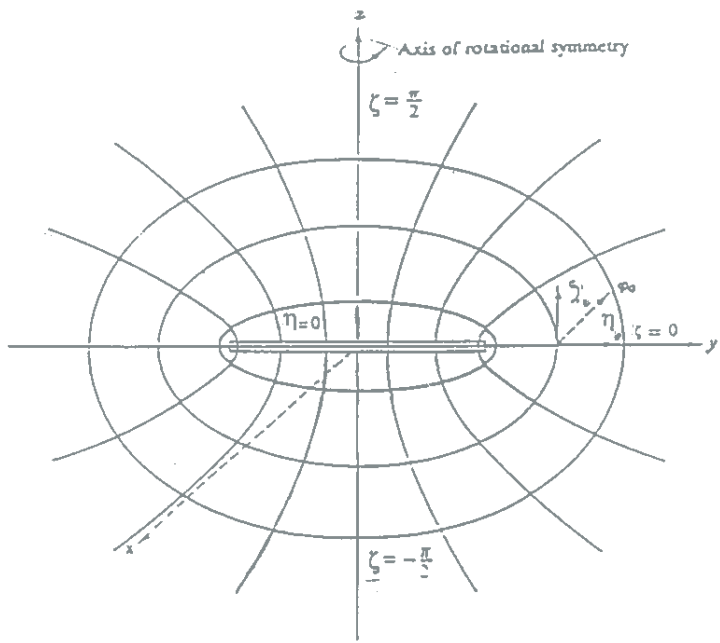

Figure 1. Oblate Spheroidal Coordinates (From Arfken. [3])

The Laplacian can then be calculated from the expression

$$\nabla^2 = \frac{1}{h_1 h_2 h_3}\left[\frac{\partial}{\partial q_1}\left(\frac{h_2 h_3}{h_1}\frac{\partial}{\partial q_1}\right) + \frac{\partial}{\partial q_2}\left(\frac{h_1 h_3}{h_2}\frac{\partial}{\partial q_2}\right) + \frac{\partial}{\partial q_3}\left(\frac{h_1 h_2}{h_3}\frac{\partial}{\partial q_3}\right)\right].$$

With the expressions for $h_1$, $h_2$, $h_3$ given above, the Laplacian becomes

$$\nabla^2 = \frac{1}{a^2(\eta^2+\zeta^2)}\left[(1+\eta^2)\frac{\partial^2}{\partial\eta^2} + 2\eta\frac{\partial}{\partial\eta} + (1-\zeta^2)\frac{\partial^2}{\partial\zeta^2} - 2\zeta\frac{\partial}{\partial\zeta}\right]$$
$$+ \left(\frac{(\eta^2+\zeta^2)}{(1+\eta^2)(1-\zeta^2)}\right)\frac{\partial^2}{\partial\varphi^2}.$$

We seek solutions to the equation

$$\nabla^2 \Psi = 0$$

Using the method of separation of variables and taking $\Psi$ to be given by

$$\Psi(\eta,\zeta,\varphi) = N(\eta)\,\Xi(\zeta)\,\Phi(\varphi)$$

The solution of Laplace's equation is then

$$\Psi(\eta,\zeta,\varphi) = \sum_j [a_j P_j(i\eta) + b_j Q_j(i\eta)][c_j P_j(\zeta) + d_j Q_j(\zeta)] e^{im\varphi}.$$

As with the solution in cylindrical coordinates, a determination of the coefficients that would produce a suitable solution is difficult since the boundary conditions are still complicated; thus we seek an alternative method of determining the potentials in the region of the pore and the membrane.

We know the Maxwell equations

$$\vec{\nabla}\cdot\vec{E} = 4\pi\rho \qquad \vec{\nabla}\cdot\vec{B} = 0$$

$$\vec{\nabla}\times\vec{E} + \frac{1}{c}\frac{\partial\vec{B}}{\partial t} = 0 \qquad \vec{\nabla}\times\vec{B} = \frac{4\pi}{c}\vec{J} + \frac{1}{c}\frac{\partial\vec{E}}{\partial t}$$

together with the implied continuity equation

$$\vec{\nabla}\cdot\vec{J} + \frac{\partial\rho}{\partial t} = 0.$$

As an alternate solution, we make use of the conditions that are to be satisfied by the fields. These conditions may be derived from Maxwells equations and can be stated as

    i.)    The tangential component of the electric field, $\vec{E}_t$ is continuous

           at an interface.

ii.) The normal component of $\vec{J}_n$ is continuous across a boundary;

In the pore region of the membrane we seek to determine the electric field, and from the field boundary conditions, the field near the tip of the membrane. The approach used is to calculate the circulation of the field around a closed loop in the region of the pore. Since we are in the D.C. limit, the field is irrotational, thus $\vec{\nabla} \times \vec{E} = 0$. In the region of the pore (region 1) the electric field may be written as $\vec{E}(\eta, \zeta)$. We choose a path of integration taken along $\eta$, and $\zeta$. From Stokes theorem we can write

$$\int_{area} (\vec{\nabla} \times \vec{E}) \cdot \vec{da} = \oint_l \vec{E} \cdot \vec{dl}$$

The field can be written in terms of its components as

$$\vec{E}(\eta, \zeta) = E_\eta(\eta, \zeta)\hat{\eta} + E_\zeta(\eta, \zeta)\hat{\zeta}.$$

This field integrated over the closed loop shown in Figure 2 gives the equation

$$\int_{\eta_1}^{\eta_2} \left[ E_\eta(\eta, \zeta_1)\hat{\eta} + E_\zeta(\eta, \zeta_1)\hat{\zeta} \right] d\vec{s}_\eta + \int_{\zeta_1}^{\zeta_2} \left[ E_\eta(\eta_2, \zeta)\hat{\eta} + E_\zeta(\eta_2, \zeta)\hat{\zeta} \right] d\vec{s}_\zeta$$
$$+ \int_{\eta_2}^{\eta_1} \left[ E_\eta(\eta, \zeta_2)\hat{\eta} + E_\zeta(\eta, \zeta_2)\hat{\zeta} \right] d\vec{s}_\eta + \int_{\zeta_2}^{\zeta_1} \left[ E_\eta(\eta_1, \zeta)\hat{\eta} + E_\zeta(\eta_1, \zeta)\hat{\zeta} \right] d\vec{s}_\zeta = 0$$

with

$$d\vec{s}_\eta = h_\eta d\eta \, \hat{\eta} = a \left( \frac{\eta^2 + \zeta^2}{1 + \eta^2} \right)^{\frac{1}{2}} d\eta \, \hat{\eta}$$

and

$$d\vec{s}_\zeta = h_\zeta d\zeta \, \hat{\zeta} = a \left( \frac{\eta^2 + \zeta^2}{1 - \zeta^2} \right)^{\frac{1}{2}} d\zeta \, \hat{\zeta}$$

combining the integrals, over d$\eta$ and d$\zeta$, and changing the limits over the second term gives

$$\int_{\eta_1}^{\eta_2} d\eta \left\{ [E_\eta(\eta,\zeta_1)] a \left( \frac{\eta^2+\zeta_1^2}{1+\eta^2} \right)^{\frac{1}{2}} - [E_\eta(\eta,\zeta_2)] a \left( \frac{\eta^2+\zeta_2^2}{1+\eta^2} \right)^{\frac{1}{2}} \right\} +$$

$$\int_{\zeta_1}^{\zeta_2} d\zeta \left\{ [E_\zeta(\eta_2,\zeta)] a \left( \frac{\eta_2^2+\zeta^2}{1-\zeta^2} \right)^{\frac{1}{2}} - [E_\zeta(\eta_1,\zeta)] a \left( \frac{\eta_1^2+\zeta^2}{1-\zeta^2} \right)^{\frac{1}{2}} \right\} = 0.$$

Since the integrals, for the different terms, are over arbitrary limits and the sum of the terms is zero, we can separately evaluate each term to zero. Thus we can write

$$\int_{\eta_1}^{\eta_2} d\eta \, a \left[ E_\eta(\eta,\zeta_1) \left( \frac{\eta^2+\zeta_1^2}{1+\eta^2} \right)^{\frac{1}{2}} - E_\eta(\eta,\zeta_2) \left( \frac{\eta^2+\zeta_2^2}{1+\eta^2} \right)^{\frac{1}{2}} \right] = 0$$

and

$$\int_{\zeta_1}^{\zeta_2} d\zeta \, a \left[ E_\zeta(\eta_2,\zeta) \left( \frac{\eta_2^2+\zeta^2}{1-\zeta^2} \right)^{\frac{1}{2}} - E_\zeta(\eta_1,\zeta) \left( \frac{\eta_1^2+\zeta^2}{1-\zeta^2} \right)^{\frac{1}{2}} \right] = 0.$$

Since the limits of the integrals are arbitrary and the terms are evaluated at different values of the parameter, $\zeta$ We can then write $E_\eta(\eta,\zeta)$ as

$$E_\eta(\eta,\zeta) = \frac{f(\eta)}{\left(\eta^2+\zeta^2\right)^{\frac{1}{2}}}.$$

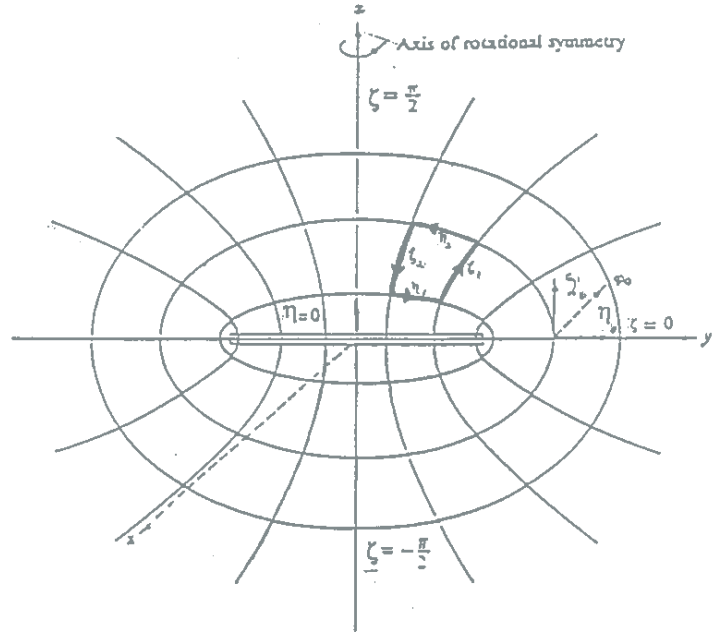

Figure 2. Line Integral Paths

a similar treatment of the $\zeta$ component of the field gives

$$E_\zeta(\eta,\zeta) = \frac{g(\zeta)}{\left(\eta^2 + \zeta^2\right)^{\frac{1}{2}}}.$$

Considering the current flow between contour lines as shown in Figure 3. From the continuity equation,

$$\vec{\nabla} \cdot \vec{J} + \frac{\partial \rho}{\partial t} = 0;$$

for steady state conditions

$$\frac{\partial \rho}{\partial t} = 0$$

and thus the continuity equation becomes

$$\vec{\nabla} \cdot \vec{J} = 0 \implies \int_a \vec{J} \cdot \hat{n}\, d\vec{a} = 0$$

where we have used the divergence theorem to obtain the last form.

We therefore must calculate the above integral which, after substituting for $\vec{J}$ and $d\vec{a}$, yields the expression that must be evaluated as

$$\int \sigma \, E(\eta,\zeta) \, h_\zeta \, h_\varphi \, d\zeta \, d\varphi$$

$$= \int_{area} \sigma \, E(\eta,\zeta) \, a^2 \frac{(\eta^2+\zeta^2)^{\frac{1}{2}}}{(1-\zeta^2)^{\frac{1}{2}}} (1+\eta^2)^{\frac{1}{2}} (1-\zeta^2)^{\frac{1}{2}} \, d\zeta \, d\varphi$$

Choosing the areas as shown in Figure 3, this integral can be written as

$$\int_{\zeta_1}^{\zeta_2} \int_{\varphi_1}^{\varphi_2} \sigma_p \, E_\eta(\eta_1,\zeta) a^2 \left( \frac{\eta_1^2+\zeta^2}{1-\zeta^2} \right)^{\frac{1}{2}} (1+\eta_1^2)^{\frac{1}{2}} (1-\zeta^2)^{\frac{1}{2}} d\zeta \, d\varphi =$$

$$\int_{\zeta_1}^{\zeta_2} \int_{\varphi_1}^{\varphi_2} \sigma_p \, E_\eta(\eta_2,\zeta) a^2 \left( \frac{\eta_2^2+\zeta^2}{1-\zeta^2} \right)^{\frac{1}{2}} (1+\eta_2^2)^{\frac{1}{2}} (1-\zeta^2)^{\frac{1}{2}} d\zeta \, d\varphi$$

This equation may be simplified to give

$$E_\eta(\eta_1,\zeta)(\eta_1^2+\zeta^2)^{\frac{1}{2}}(1+\eta_1^2)^{\frac{1}{2}} = E_\eta(\eta_2,\zeta)(\eta_2^2+\zeta^2)^{\frac{1}{2}}(1+\eta_2^2)^{\frac{1}{2}}$$

taking $E_\eta$ from previous

$$E_\eta(\eta,\zeta) = \frac{f(\eta)}{(\eta^2+\zeta^2)^{\frac{1}{2}}}$$

and substituting into the expression above gives

$$\frac{f(\eta_1)}{(\eta_1^2+\zeta^2)^{\frac{1}{2}}}(\eta_1^2+\zeta^2)^{\frac{1}{2}}(1+\eta_1^2)^{\frac{1}{2}} = \frac{f(\eta_2)}{(\eta_2^2+\zeta^2)^{\frac{1}{2}}}(\eta_2^2+\zeta^2)^{\frac{1}{2}}(1+\eta_2^2)^{\frac{1}{2}}$$

which simplifies to

$$f(\eta_1)(1+\eta_1^2)^{\frac{1}{2}} = f(\eta_2)(1+\eta_2^2)^{\frac{1}{2}}.$$

Since the two expressions are equal for arbitrary values of $\eta_1$ and $\eta_2$, we can write

$$f(\eta) = \frac{E_\eta^0}{(1+\eta^2)^{\frac{1}{2}}}$$

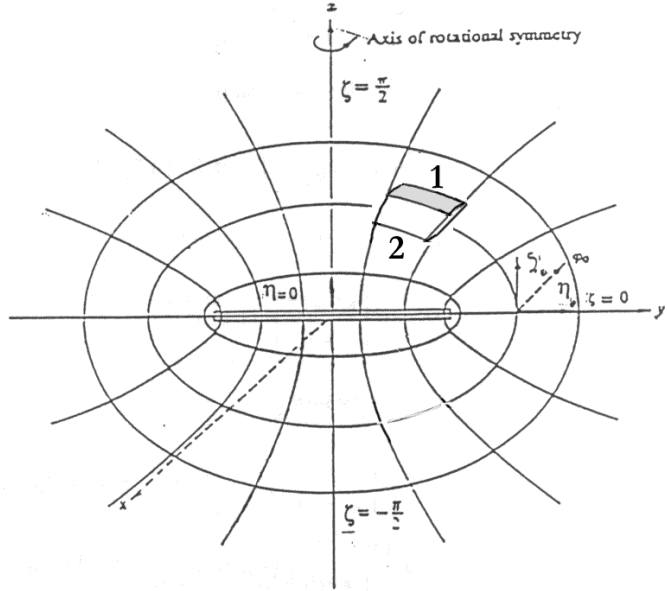

Figure 3. Schematic Showing Areas for Current Flow

so that

$$E_\eta(\eta, \zeta) = \frac{E_\eta^0}{(1+\eta^2)^{\frac{1}{2}}(\eta^2 + \zeta^2)^{\frac{1}{2}}}.$$

This procedure may be repeated for the $\zeta$ component of the electric field, giving

$$\iint \sigma_1 E_\zeta(\eta, \zeta_1) da_{\eta\varphi} = \iint \sigma_2 E_\zeta(\eta, \zeta_2) da_{\eta\varphi}$$

where

$$da_{\eta\varphi} = h_\eta d\eta\, h_\varphi d\varphi = a^2 \left(\frac{\eta^2 + \zeta^2}{1+\eta^2}\right)^{\frac{1}{2}} (1+\eta^2)^{\frac{1}{2}}(1-\zeta^2)^{\frac{1}{2}} d\eta\, d\varphi$$

so that

$$g(\zeta) = \frac{E_\zeta^0}{\left(1-\zeta^2\right)^{\frac{1}{2}}}$$

and

$$E_\zeta(\eta,\zeta) = \frac{E_\zeta^0}{\left(1-\zeta^2\right)^{\frac{1}{2}}\left(\eta^2+\zeta^2\right)^{\frac{1}{2}}}$$

The z axis corresponds to the value of $\zeta = 1$ and is included in the region of the pore in the membrane. From the expression for $E_\zeta(\eta,\zeta)$ above points on the z axis would have an electric field with a magnitude which is undefined; thus we take $E_\zeta^0 = 0$. We see, therefore, that in the pore, the field is entirely in the $\eta$ direction and is given by the expression,

$$E_\eta(\eta,\zeta) = \frac{E_\eta^0}{\left(1+\eta^2\right)^{\frac{1}{2}}\left(\eta^2+\zeta^2\right)^{\frac{1}{2}}}.$$

where the parameter $E_0^\eta$ depends only on the external applied field.

**Conclusion**

We have derived the form of the electric field in the region of a pore formed in the membrane. Due to problems of directly obtaining the solution from Laplaces equation, we have adopted a different approach based on properties of the field. This approach allows us to derive a functional form for the electric field in the region of the pore.

**References**

Abidor, I. G., Arakelyan, V. B., Chernomodik, L .V., Chizmadzhev, Y. A., Pastushenko, V. F., and Tarasevich, M. R. Electrical breakdown of BLM: Main experimental facts and their qualitative discussion. *Bioelectrochem. Bioenerg.*, **6**, 1979, 37-52.

Arfken, George. Mathematical Methods for Physicists 2nd. Edition. Academic Press, New York and London. 1970.

Barnett, Weaver Electroporation: A Unified quantitative theory of reversible electrical breakdown and rupture. *Bioelectrochem. Bioenerg.*, **25**, 1991, 163-182.

Chernomordik, L. V., S. I. Sukharev, I. G. Abidor and Yu A. Chizmadzhev. Breakdown of Lipid Bilayer Membranes in an Electric Field. *Biochimica et Biophysica Acta*, **736**, 1983, 203-213.

Coster H. G. L., and Zimmermann, U. The mechanism of electrical breakdown in the membranes of *Valonia Utricularis*. *J .Membrane Biol.*, **22**, 1975, 73-90.

Crowley, J. M. Electrical breakdown of bimolecular lipid membranes as an electro-mechanical instability. *Biophysical Journal*, **13**, 1973, 711-724.

Drago, D. P. And S. Ridella. Evaluation of electrical fields inside a biological structure. *Br. J. Cancer*, **45**, Suppl. V, 1982, 215-219.

Gruen. David, W. R., and Joe Wolfe. Lateral Tensions and Pressures in Membranes and Lipid Monolayers. *Biochimica et Biophysica Acta*, **688**, 1982, 572-580.

Jackson, John David . Classical Electrodynamics. Second Edition.   John Wiley and Sons, Inc., New York London Sydney Toronto. 1975.

Kinosita, K., Jr. And Tsong, T. Y. Formation and resealing of pores of controlled sizes in human erythrocyte membrane. *Nature*, **268**,  1977, 438-441.

Marchesi, M.,  M. Parodi. Study of the electrical field inside biological structures*.  Med. & Biol. Eng. & Comput.*, **20**, 1982, 608-612.